\documentclass[aps,prl,twocolumn,groupedaddress,floatfix]{revtex4}
\usepackage{graphicx}

\begin{document}
\author{Alejandro Lopez-Bezanilla}
\email{alejandrolb@gmail.com}
\affiliation{Oak Ridge National Laboratory, One Bethel Valley Road, Oak Ridge, Tennessee, 37831-6493, USA}

\title{Electronic Transport Properties of Chemically Modified Double-Walled Carbon Nanotubes}

\begin{abstract}
We present a study on the quantum transport properties of chemically functionalized metallic double-walled carbon nanotubes (DWNTs) with lengths reaching the micrometer scale. First-principles calculations evidence that, for coaxial tubes separated by the typical graphitic van der Waals-bond distance, the chemical modification of the outer wall with sp$^3$-type defects affects the electronic structure of both the outer and the inner tube, which reduces significantly the charge transport capability of the DWNT. For larger spacing between sidewalls, the conductivity of the outer wall decreases with increasing functional group coverage density while charge transport in the inner tube is equivalent to that of a pristine nanotube. Additionally, chemical attachment of CCl$_2$ onto the outer DWNT sidewall barely affect the conjugated $\pi$-network of the double-wall and charge transport remains in the quasi-ballistic regime. These results indicate an efficient route for tailoring electronic transport in DWNTs provided inner shell geometry and functional groups are properly chosen.
This document is the Accepted Manuscript version of a Published Work that appeared in final form in The Journal of Physical Chemistry C, copyright American Chemical Society after peer review and technical editing by the publisher. To access the final edited and published work see http://pubs.acs.org/doi/abs/10.1021/jp402355x
\end{abstract}
\maketitle

\section{Introduction}

Due to their unique physical and chemical properties, carbon nanotubes (CNTs) are exceptional nanostructures enabling the fabrication of new types of functional devices. Their quasi-one-dimensional structure, high room-temperature charge mobility and current capacity have attracted an enormous interest for their integration in high-performance integrated circuits. \cite{Derycke, ISI:000288003900008} CNTs engineering is allowing their utilization in numerous scientific and industrial applications for, e.g. chemical sensing applications. The high-sensitivity of CNTs to its environment leads to modifications on the transport properties that can further evidence the presence of ultralow concentration of analytes, such as explosives, viruses or toxics. Also, CNTs have been identified as a promising option in organic photovoltaics as an alternative to transparent conducting materials such as indium tin oxide, allowing for the long-term viability of solar-power based on abundant elements and cost-efficient materials.\cite{pelectronically2011}

Several approaches have been proposed to control and add new functionalities to the intrinsic outstanding electric properties of CNTs.\cite{ISI:000075621600011} \cite{RMProche} In this context, the covalent attachment of substituted functional phenyl groups to individual CNTs via the diazonium coupling route \cite{BahrTour} represents an efficient method to anchor more complex chemical functionality onto the external nanotube sidewall. Further attachment of photoactive molecules to the phenyl groups results in a nanotube-based device that can be sensitive to external radiation. However, addition reactions to the nanotube C-C double bonds involves the transformation of sp$^2$- into sp$^3$-hybridized carbon atoms. This transformation significantly alters the electronic properties of the hyperconjugated p$_z$-network, as evidenced by absorption spectra,\cite{Strano12092003} and damages the nanotube electronic conducting capability. Although the diazonium reaction is reversible and the functional groups can be removed from the surface, the degradation of the CNT features is inevitable as a result of the accumulation of sidewall defects. \cite{doi:10.1021/ja068320r} Noncovalent covering of nanotube surface with molecules deals with issues such as the short-term stability of the physisorbed groups, owing to the weak interaction and the low charge transfer efficiency between both systems. To circumvent such a problem, several strategies have been proposed. 

The first was pointed out by theoretical investigations\cite{PhysRevLett.97.116801} which predict that nanotube sidewall [2+1] cycloaddition of divalent groups, such as diclorocarbene (CCl$_2$), would allow for the preservation of the nanotube conjugated $\pi$-network, while providing anchoring sites for further attachment of functional molecules upon Cl substitution. In this case, divalent addition occurs when the C atom of CCl$_2$ bridges two neighboring C atoms of the tube, which entails the cleavering of the $\sigma$-bonds between the adjacent C atoms and the creation of two new $\sigma$-bond between the former and the latter. Sidewall [2+1] cycloaddition on small-diameter armchair nanotubes with C-C bond orientations perpendicular to the tube axis is the only possibility for the host atoms to reorient the $\sigma$-bond to the guest C atom, and leave the nanotube $\pi$-network and electronic properties almost unaltered.\cite{ISI:000276179400007} For large diameter armchair CNTs and other C-C bond orientations, as in semiconducting zigzag nanotubes,\cite{ ISI:000276179400007} a transport gap develops as a consequence of the quasi-bound states introduced by the sp$^3$-type defects, yielding to electronic backscattering events that dramatically decrease the transport capability of the tubes. \cite{ISI:000312301800016,ISI:000307676600072} 

An alternative consists in taking advantage of the coaxial geometry of double-walled carbon nanotubes (DWNTs) in which the outer wall is conveniently functionalized while the inner wall is protected from chemical agents and more enable to convey electrical current. While the  modification of single-walled CNTs (SWNTs) with chemical agents and its consequences on their transport properties have been extensively studied, both theoretically and experimentally, detailed investigations on the effect of outer-sidewall grafted coatings on the electronic and transport properties of a double-walled system still remain a challenging task. For instance, it is not clear if the new electronic states introduced by the functional groups upon rehybridization of outer wall C atoms can affect the conducting channels of the inner nanotube. Indeed, Raman measurements in DWNTs reported by Huang et al. evidenced a possible wall-to-wall interaction due to the outer-sidewall chemical modification.\cite{Huang} Conductance changes in SWNTs upon geometric, structural or chemical modification have been the object of numerous studies over the last two decades, although very little attention has been paid to the modification of conducting properties of multi-walled CNTs from realistic models. 

In this paper we present a first-principles computational study of the electronic structure and the charge transport properties of DWNTs as a function of the inter-wall separation and the random distribution of phenyl and diclorocarbene functional groups attached to the outer tube sidewall. Using a combination of density-functional-theory-(DFT)-based calculations and simulations of mesoscopic electronic-transport, we unveil a broad spectrum of transport features in functionalized DWNTs at the micrometer scale, ranging from the quasi-ballistic to the insulating regime. For the typical van der Waals inter-wall separation distance of $\sim$3.3 \AA, phenyl functionalization of the outer wall is observed to alter the electronic properties of both the inner and the outer tube. This reduces the electronic-transport capability of the DWNT, leading to a strongly localized transport regime. For larger inter-wall distances, the conductance of the outer wall tube decays rapidly with increasing number of sp$^3$-type defects, whereas the inner tube conductance ability remains intact. In contrast, outer tube wall modification with CCl$_2$ functional groups were shown to preserve good conduction ability for both inner and outer tubes for an inter-wall separation of $\sim$3.3 \AA, preserving the original ballistic conduction of pristine DWNTs. The study is performed for realistic length-scales of up to 4 $\mu$m-long functionalized DWNTs. Although this transport methodology neglects physical effects that will take place in an out-of-equilibrium situation such as the electron-phonon scattering, the influence of solvants or a strong applied bias, it allowed us to determine the intrinsic effects of chemical surface defects on the transport features of DWNTs.

 \section{Methodology}
 
\subsection{Ab initio scheme}
The geometry optimizations and electronic structure calculations were performed with the SIESTA code.\cite{PhysRevB.53.R10441,0953-8984-14-11-302} A double-$\zeta$ basis set within the local density approximation (LDA) approach for the exchange-correlation functional was used. Although van der Waals functionals are known to describe accurately the interaction between the concentric layers of pristine double-walled carbon nanotubes, LDA was considered as a suitable functional since the dominating effects here are the new states coupling the concentric tubes upon chemical modifications. Also LDA allows us for the required level of accuracy in the description of large unit cells of up to 1010 atoms of chemically modified coaxial nanotubes at an affordable computational cost, which is $\approx$10 times less expensive than van der Waals functionals.\cite{PhysRevLett.103.096102} Carbon nanotubes were modeled within a supercell large enough to avoid interactions between neighboring cells. Atomic positions were relaxed with a force tolerance of 0.02 eV/\AA. The integration over the Brillouin zone was performed using a Monkhorst sampling of 1x1x4 k-points for 13-primitive armchair unit cell long tubes with functional groups. The radial extension of the orbitals had a finite range with a kinetic energy cutoff of 50 meV. The numerical integrals were computed on a real space grid with an equivalent cutoff of 300 Ry.
 
\subsection{Electronic-transport approach}
To determine the electronic-transport properties of modified DWNTs we resort to the Landauer-B\"uttiker formulation of the conductance, which is particularly suitable to study the electron motion along a one-dimensional device channel in between two semi-infinite electrodes. The scattering region where charge carriers can be backscattered during their propagation is a defective phase-coherent multi-mode channel (functionalized DWNT) in between two semi-infinite electrodes (pristine DWNTs), which are in thermodynamical equilibrium with infinitely larger electron reservoirs. The computational strategy used for the description of the electronic-transport properties is based on a well-stablished scheme as discussed in prior works \cite{Biel, swnt}. A set of first-principles calculations are first performed to obtain the {\it ab initio} Hamiltonians (H) and overlap (S) matrices associated with DWNT sections whose outer wall is modified by functional groups. The atomic-like basis set utilized by the SIESTA code allows us to obtain a description of the DWNT-functional group coupling with relatively small and manageable sparse Hamiltonian matrices. The functionalized DWNT unit cells are long enough so that the nanotube-extremes are converged to the clean system. Thus functionalized and clean sections of DWNTs can be matched to construct long systems with perfect contact areas between the building blocks. Pieces of both modified and pristine sections are assembled in a random fashion to mimic rotational and translational disorder. Standard techniques to calculate the Green functions associated with the sparse Hamiltonians and overlap matrices are further used to include recursively the contribution of the sections, within a O(N) scheme with respect to the tube length. Energetic descriptions of micrometer-long tubes can therefore be attained within the accuracy of first-principles calculations. To evaluate the conductance $G$ of the system we adopt the standard Green function formalism which, at quasi-equilibrium conditions, is $G(E)={G_0}\sum_{n}T_n(E)$, where $G_0=e^{2}/h$ is the quantum of conductance. The transmission coefficients $T_n(E)$ are calculated by evaluating the retarded (advanced) Green functions of the system:
\begin{equation}
\mathcal{G}^{\pm}(E)=\{E S-H-\Sigma^{\pm}_{L}(E)-\Sigma^{\pm}_{R}(E)\}^{-1}
\end{equation}
where $\Sigma^{\pm}_{L(R)}(E)$ are the self-energies which describe the coupling of the channel to the left (right) electrodes. These quantities are related to the transmission factor by the Caroli formula \cite{0022-3719-4-8-018}:
\begin{equation}
T(E)=tr\{\Gamma_{L}(E) \mathcal{G}^{+}(E) \Gamma_{R}(E) \mathcal{G}^{-}(E)\}
\end{equation}
with 
$\Gamma_{L(R)}(E)=i\{\Sigma^{+}_{L(R)}(E)-\Sigma^{-}_{L(R)}(E)\}$. $tr$ stands for the trace of the corresponding operator. 

The transmission coefficients $T_n(E)$ for a given channel $n$ give the probability of an electron to be transmitted at energy $E$ from the source to the drain electrode. For a pristine nanotube, $T_n(E)$ assumes integer values corresponding to the total number of open propagating modes at the energy $E$. For an isolated armchair CNTs, $T_n(E)=2$ near the Fermi level (first plateau) and 4 for the second plateaus. Given that tube-tube interaction is weak, the conductance of a pristine DWNT in the ballistic conduction regime is given as the sum of all conducting channels of each tube at a given energy.

\section{Results and discussion}

\begin{figure}[htp]
 \centering
 \includegraphics[width=0.45 \textwidth]{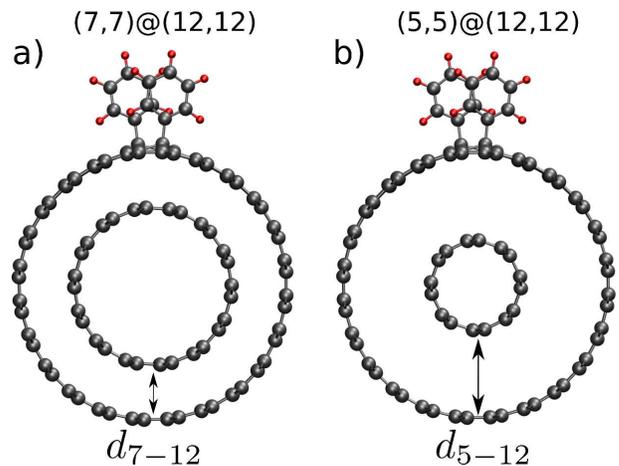}
 \caption{Atomic structures of the paired phenyl functionalized double-walled carbon nanotubes (DWNT) studied here. a) (7,7)@(12,12) DWNT with two phenyl functional groups covalently attached to the outer-sidewall. The inter-wall distance $d_{7-12}$ is $\sim$3.3 \AA. b) (5,5)@(12,12) DWNT with two phenyl groups attached at the same C atoms of the previous structure. The inter-wall distance $d_{5-12}$ is $\sim$5.5 \AA.}
 \label{FIG1}
\end{figure}

In the following, we refer to two armchair concentric CNTs as (n,n)@(m,m), where n stands for the index of the smaller CNT. Any two CNTs following the rule m=n+5 have an inter-wall spacing of $\sim$3.3 \AA, close to that of the van der Waals-bonded planar graphite sheets. Both tubes are assumed to be rigid cylinders with parallel axes, and the double-wall tubular geometry is defined by the inter-wall separation. To construct a minimal functionalized system, concentric armchair DWNT unit cells were repeated 13 times along the z-axis such that the geometric and energetic perturbation caused by a functional group grafted on the outer-sidewall vanish at the edges of the supercell. The entire system of up to 1010 atoms is fully relaxed. Here we focus first on a paired phenyl functionalized (7,7)@(12,12) and (5,5)@(12,12) DWNT, as schematically represented in Figure \ref{FIG1}, where two phenyls are grafted at third-nearest neighboring positions. This 1,4-geometry of phenyls rings is considered as the most likely to be found in graphitic systems \cite{ISI:000242297500004} for several reasons: firstly, the removal of a p$_z$-orbital of the conjugated network as a consequence of a phenyl attachment is known to create an unstable radical in the vicinity of the grafting site which enhances the reactivity of C atoms at an odd number of bonds away from it \cite{ISI:000242297500004,Schmidt}. Secondly, the 1,4-configuration is more stable than the 1,2-configuration due to the steric repulsion between the face-to-face phenyl groups. Thirdly, isolated phenyls have been predicted to desorb at room temperature. \cite{Margine} 

\begin{figure*}[htp]
 \centering
 \includegraphics[width=0.95 \textwidth]{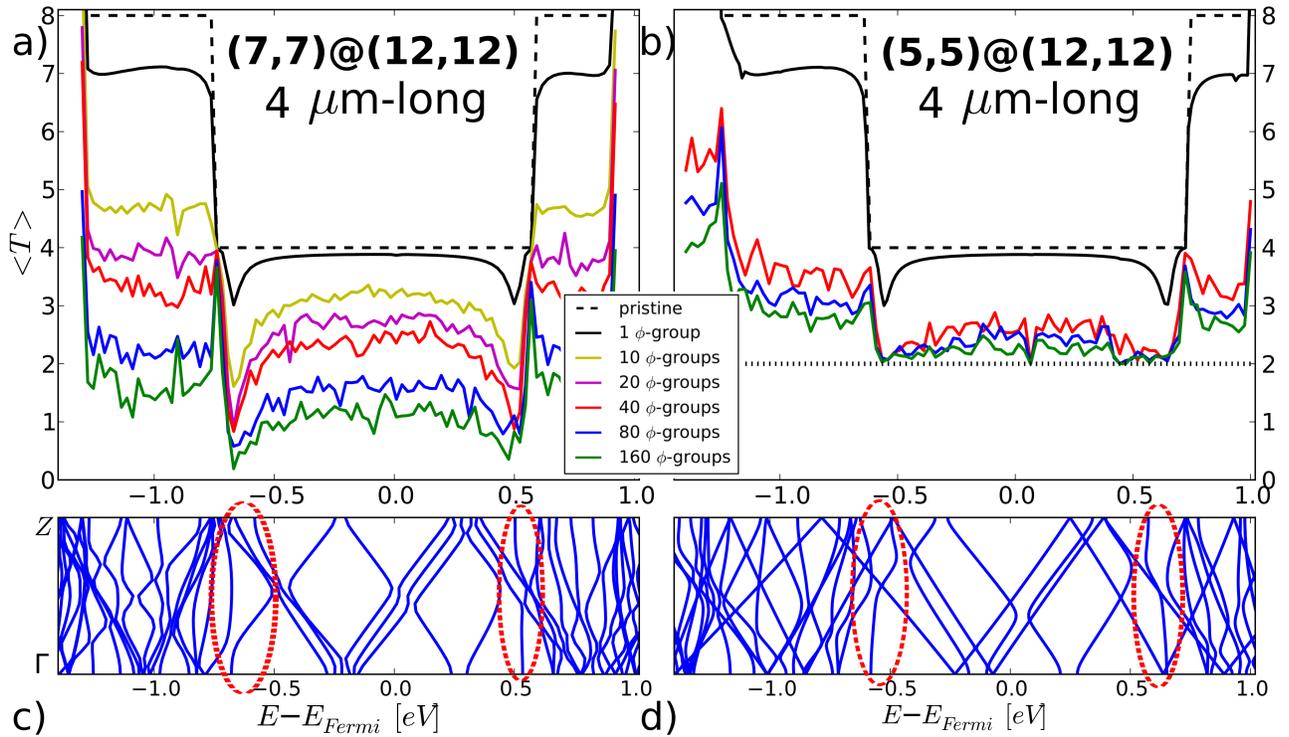}
 \caption{ a) Averaged transmission coefficients for 4 $\mu$m-long (7,7)@(12,12) DWNTs whose outer-sidewall is functionalized with paired phenyl groups. The effect of a single pair of phenyl groups is shown, as well as the decreasing  transmission for an increasing number of groups on the outer surface. b) Same as in a) but for the (5,5)@(12,12) DWNT. c) Electronic band diagram of a 13-unit cell (7,7)@(12,12) DWNT functionalized with paired phenyls. Red ovals indicate the position in energy of the antiresonant states responsible of the dip of transmission shown in a) for a single group. d) Same as in c) but for the (5,5)@(12,12) DWNT. The transmission coefficient for a pristine DWNT is in dashed lines, and curves have been averaged over 40 different random configurations. In b), horizontal dotted line indicates T(E)=2}
 \label{FIG2}
\end{figure*}

\begin{figure*}[htp]
 \centering
 \includegraphics[width=0.93 \textwidth]{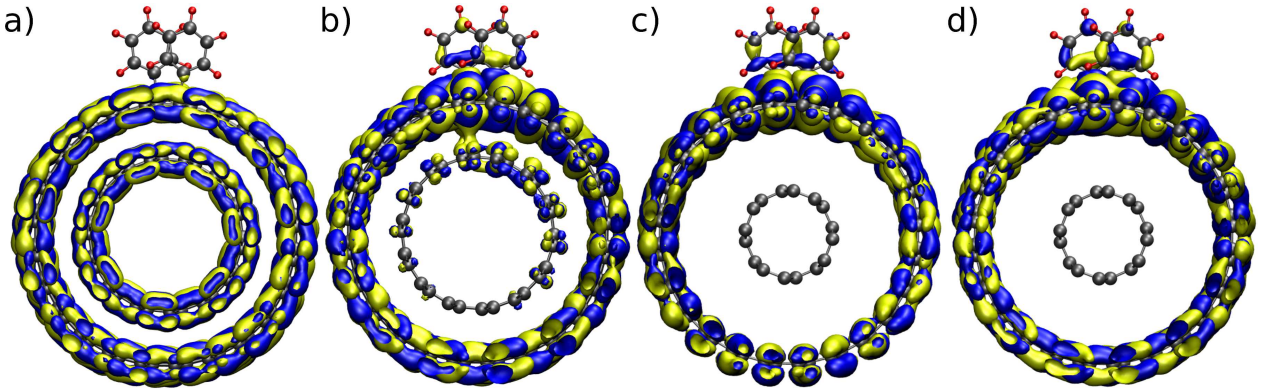}
 \caption{ Electronic wavefunction plots at the $\Gamma$-point of the quasi-bound states at which the antiresonances are found in Figure \ref{FIG2}. For a (7,7)@(12,12) DWNT the two phenyl groups on the outer surface interact with both outer and inner tubes, creating two localized states in the hole a) and electron b) bands that extend to both tubes. On the contrary, for a (5,5)@(12,12) phenyl functionalization only affects the outer tube, leaving the inner tube unaffected. The wavefunction for the quasi-bound state for holes is in c), and for electrons in d), showing that in either case the resonant state extends over the inner tube.}
 \label{FIG3}
\end{figure*}

The averaged transmission profiles of (7,7)@(12,12) and (5,5)@(12,12) DWNTs with an increasing number of phenyl groups are plotted in Figure \ref{FIG2}-a) and b), respectively. As a reference for the next results, the transmission profiles of pristine DWNTs are plotted with dashed lines. It is of special interest to analyze the effect of two sp$^3$ rehybridizations introduced by a group of paired phenyls in the transmission of a pristine DWNTs. Such a type of functionalization provokes a conductance drop for all energy values, with an occurrence of two symmetric dips in the first plateau, (at -0.67 eV and 0.50 eV for the (7,7)@(12,12), and at -0.56 eV and 0.64 eV for the (5,5)@(12,12)). Examining the energy band diagrams of the corresponding periodic systems (Figure \ref{FIG2}-c) and d)), one notices the presence of two low-dispersive states (pointed out by red ovals) at the same energy values of those of the two conductance dips. The resonant backscattering of these quasi-bound states reduces the conductance in one transmission channel. There is, nevertheless, a significant difference between the low-dispersive bands of both diagrams, namely, the markedly mixing of electronic states in the case of (7,7)@(12,12) DWNT that does not occur in the (5,5)@(12,12), which points out to different degrees of hybridization of the phenyl molecular states with the $\pi$-conjugated network of the two coaxial nanotubes. The formation of two sp$^3$ DWNT-phenyls bonds involves a change from a trigonal-planar local bonding geometry to a tetrahedral geometry, and the combination of the phenyl electronic states with those of the outer tube. In the case of the (7,7)@(12,12) DWNT, due to the short inter-wall spacing, this leads to an additional hybridization with the inner nanotube electronic states, resulting in the mixing of the three nanostructure bands, as observed in Figure \ref{FIG2}-c). The hybridization occurs not with particular states of the inner tube but with most $\pi$ and $\sigma$ orbitals, as evidenced by several band gaps opening for a large energy range, affecting the transport ability of the quantum modes. In Figure \ref{FIG3}-a) and b) the electronic wavefunctions of the quasi-bound states show that both outer and inner tubes participate in the hybridrization. Differently for the (5,5)@(12,12) DWNT, the large wall-to-wall distance implies that solely outer-sidewall C atoms participate in the bonding with the phenyl groups. Thus, the inner tube preserves its metallic character with a largely dispersive state, and its $\pi$ character. As evidenced in Figure \ref{FIG3}-c) and d), the two quasi-bound states spread uniquely over the outer tube around the phenyl groups.

To analyze how the lack of translational invariance dictates the electronic-transport regime, we have conducted a mesoscopic study for several tubes with lengths of up to 4 $\mu$m and random grafting of functional groups. Conductances were averaged over 40 disordered configurations. For a large number of groups grafted onto the surface of a (7,7)@(12,12) DWNT, both translational and rotational symmetries are broken, and it is thus expected that multiple scattering phenomena on both sidewalls eventually reduces the conductance of the system down to the localized regime. This is indeed observed in Figure \ref{FIG2}-a) when an increasing number of paired phenyls from 10 up to 160 are randomly distributed on the outer-sidewall. The transmission drop of one single group is amplified when adding a much larger number of sp$^3$ scatters, which enhances the effect of disorder and the quantum interferences between the quasi-bound states. For 160 paired phenyls the averaged transmission coefficient is below 1, indicating that the system has entered the localized quantum transport regime. A different trend is observed in the case of functionalized (5,5)@(12,12) DWNT, for which the transmission drop does not decrease below the limit of T(E)=2 in the first plateau. As a result, for the same number of grafted groups, only the conducting channels of the outer nanotube are succesively closed as the number of sp$^3$ defects increases, whereas the conducting ability of the metallic inner tube remains unaffected. These results are in resonance with the experimental investigations reported by Bouilly et al. \cite{doi:10.1021/nn201024u} on electrical measurements on DWNTs whose outer wall were covalently functionalized with aryldiazonium salts. They concluded that the electrical current in a functionalized DWNT is comparable to the current carried by a pristine single-walled CNT, i.e. the inner tube is electrically active upon outer-sidewall modification. The transmission patterns of Figure \ref{FIG2}-a) are actually similar to the ones already reported for SWNT with similar functionalization in ref. \cite{swnt}, suggesting that an outer wall sp$^3$-type defected DWNT with short inter-wall spacing behaves as a large diameter SWNT in transporting electric current. Similarly, coaxial DWNTs with large inter-wall separation behave as two independent nanotubes.

\begin{figure}[htp]
 \centering
 \includegraphics[width=0.45 \textwidth]{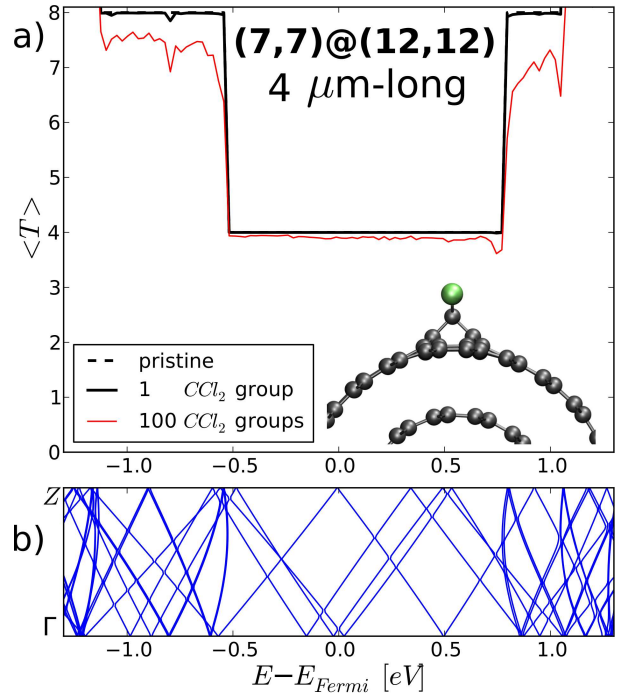}
 \caption{  Averaged transmission coefficients for 4 $\mu$m-long (7,7)@(12,12) DWNTs whose outer-sidewall is functionalized with 100 CCl$_2$ groups. The transmission curve for a single group is very similar to that of the pristine DWNT due to the weak modification on the electronic properties of the DWNT introduced by the [2+1] cycloaddition. Unlike band diagrams of phenyl functionalized DWNTs in Figure \ref{FIG2}, the electronic band diagram in b) does not exhibit any localized state. Inset in a): a CCl$_2$ group bridging two C atoms of the outer-sidewall.}
 \label{FIG44}
\end{figure}

In contrast, the cycloaddition of diclorocarbene groups yields a small downscaling of the transmission with respect to the coverage density and inter-wall separation of concentric tubes. The most stable orientation of CCl$_2$ groups on the outer-sidewall has been considered, where the C atom of the functional group is simultaneously attached to two neighboring nanotube C atoms in the same circumferential plane (see inset of Figure \ref{FIG44}). The random distribution of up to 100 groups on 4 $\mu$m-long (7,7)@(12,12) DWNTs barely modifies the transmission profile of the pristine system, and keeps the system in the quasi-ballistic regime. The much weaker change of conductance  demonstrates the noninvasive effects of [2+1] cycloadditions in both the disruption of the $\pi$-conjugated network of the concentric nanotubes and their transport abilities, which is decisive in further utilization of hybrid nanotubes.

\section{Conclusions}

In conclusion, the effects of the chemical modification of DWNT outer-sidewall with two types of chemical functionalization have been examined. By means of first-principles descriptions and electronic transport calculations, various conduction regimes have been obtained in micrometer-long modified DWNTs depending on the energy of incoming charges, as well as on the coverage density of chemical addends, and the inter-wall separation. For short inter-wall separation, the weak wall-to-wall coupling of pristine DWNTs is significantly altered by phenyl functionalization of outer-sidewall, yielding strongly diffusive regime. In contrast, for larger inter-wall spacing, this sp$^3$-type functionalization only affects the transport ability of the outer tube whereas the inner tube remains in the quasi-ballistic regime. Different from the case of phenyl groups, [2+1] cycloaddition of diclorocarbene groups to the outer-sidewall yields a small downscaling of the conductance for all energies, regardless the coverage rate and the inter-wall spacing. Our study quantifies the limits of the sp$^3$-type functionalization in DWNT with short inter-wall separation and suggests that DWNTs with large inter-wall spacing or sp$^2$-type functionalization should be targeted to engineer efficient novel device functionalities. These findings are consistent with previously reported experimental studies and provide further understanding of the quantum transport properties of chemically modified DWNTs, predicting that inter-wall spacing represents a key factor to consider when assembling novel electronic devices based on DWNT composites with well-defined chemical functions.

\section{Acknowledgements}
This research used resources of the National Center for Computational Sciences at Oak Ridge National Laboratory, which is supported by the Office of Science of the U.S. Department of Energy under Contract No. DE-AC05-00OR22725. I am also grateful for the support from the Center for Nanophase Materials Sciences (CNMS), sponsored at Oak Ridge National Laboratory by the Division of Scientific User Facilities, U.S. Department of Energy.

%
%


\providecommand*\mcitethebibliography{\thebibliography}
\csname @ifundefined\endcsname{endmcitethebibliography}
  {\let\endmcitethebibliography\endthebibliography}{}

\end{document}